\documentclass{aa}  
\usepackage{hyperref}
\hypersetup{colorlinks=true,
            allcolors=blue}
\usepackage{graphicx}
\usepackage{booktabs}
\usepackage{wasysym} 
\usepackage{ragged2e} 

\newcommand{\kms}[0]{km\,s$^{-1}$}
\newcommand{\gs}[0]{g\,s$^{-1}$}
\newcommand{\hei}[0]{\ion{He}{i}\,1083\,nm }
\usepackage{txfonts}
\begin{document} 

%------ TITLE, AUTHORS, ABSTRACT ------%
\title{Cold dayside winds shape large leading streams in evaporating exoplanet atmospheres}
\titlerunning{Cold dayside winds shape large leading streams}

\author{F. Nail \inst{1},
        M. MacLeod \inst{2},
        A. Oklop\v{c}i\'{c} \inst{1}, M. Gully-Santiago \inst{3}, C.V. Morley \inst{3}, Z. Zhang \inst{3,4}}
    
\authorrunning{F. Nail et al.}
\institute{Anton Pannekoek Institute for Astronomy, University of Amsterdam, 1090 GE Amsterdam, Netherlands
\and 
Center for Astrophysics, Harvard \& Smithsonian 60 Garden Street, MS-16, Cambridge, MA 02138, USA
\and 
Department of Astronomy, The University of Texas at Austin, 2515 Speedway, Austin, TX 78712, USA
\and
Department of Astronomy and Astrophysics, University of California, Santa Cruz, Santa Cruz, CA 95064, USA
}
\date{Published in 
    A\&A, Volume 695, id.A186, March (2025).}
 
\abstract{
   Recent observations of planetary atmospheres in HAT-P-32~b and HAT-P-67~b reveal extensive outflows reaching up to hundreds of planetary radii. The helium 1083 nm light curves for these planets, captured across their full orbits, show notable asymmetries: both planets display more pronounced pre-transit than post-transit absorptions, with HAT-P-67~b being the more extreme case. Using 3D hydrodynamic simulations, we identified the key factors influencing the formation of a dense leading outflow stream and characterized its morphology. Our models suggest that such a geometry of escaped material is caused by a relatively cold outflow with a high mass-loss rate, launched preferentially from the planet's dayside. From the simulations we calculated synthetic \hei spectra that show large absorption depths and irregular line profiles due to complex gas kinematics. We find that the measurements of the \hei equivalent width and the velocity shift relative to the planet's rest frame, observed over a significant portion of the planet's orbital phase, can provide important constraints on the outflow properties and its interaction with the stellar wind.}

\maketitle

%------ INTRODUCTION ------%

\section{Introduction}

%% INTRO
The study of atmospheric escape in planets is crucial for understanding the mass loss and long-term evolution of planetary atmospheres \citep[e.g.,][]{owen_atmospheric_2019}. Spectroscopic observations, particularly in the \hei line, have become essential tools for tracing the kinematics and structures of these outflows \citep[e.g.,][]{oklopcic_new_2018, spake_helium_2018, nortmann_ground-based_2018, allart_spectrally_2018, salz_detection_2018}. Recent \hei transit observations conducted with the Habitable-zone Planet Finder Spectrograph (HPF) on the 10-meter Hobby-Eberly Telescope (HET) of HAT-P-32~b \citep{zhang_giant_2023} and HAT-P-67~b \citep{gully-santiago_large_2024} have revealed planetary outflows extending over hundreds of planetary radii. HAT-P-67~b is particularly intriguing due to its exceptionally extended pre-transit outflow, reaching orbital phases as early as $\varphi = -0.3$, which is followed by a sharp drop immediately after the optical transit. The helium light curve of HAT-P-32~b also shows a notable asymmetry, with a stronger pre-transit phase that lasts an hour longer than the post-transit phase. Both planets display remarkably high helium excess depths: 8.2\% for HAT-P-32~b and 10\% for HAT-P-67~b. These predominantly leading, high-density outflows differ from the more typical scenario in which the outflow is more extended on the trailing side, behind the planet \citep[e.g.,][]{ehrenreich_giant_2015, nortmann_ground-based_2018, guilluy_gaps_2020, spake_posttransit_2021, ben-jaffel_signatures_2022, Tyler2024}. This unusual morphology motivated us to investigate what outflow properties can cause such a configuration using 3D hydrodynamic simulations.

%% HAT-P-67b 
The hot Saturn HAT-P-67~b \citep{zhou_hat-p-67b_2017, gully-santiago_large_2024} orbits an evolved F5 subgiant (effective temperature $T_{\rm eff} = 6406$~K) at a distance of $a = 0.062$~au, with an orbital period of $P_{\rm orb} = 4.81$ days and a transit duration of 7.0 hours. The planet's proximity to its host star, coupled with the stellar evolution through the subgiant phase and increased luminosity, results in notably high insolation for HAT-P-67~b. The planet has an equilibrium temperature of $T_{\rm eq} = 1903$~K,  a reported upper mass limit of $M_p < 0.59 M_J$, and a radius of $R_p = 2.17 R_{\rm J}$, making it inflated with a low surface gravity of $\log g_p < 2.3$~dex. HAT-P-67~b is assumed to be on a circular orbit, and Doppler tomographic observations have shown that the planet's orbital plane is closely aligned with the equatorial plane of its host star \citep{zhou_hat-p-67b_2017, sicilia2024gapsprogrammetngxxx}. The HET observations suggest a high planetary mass-loss rate on the order of $ 10^{13}$~g~s$^{-1}$ based on the comparison with spherically symmetric (1D) Parker wind models \citep{gully-santiago_large_2024}. Additional high-resolution transit data were obtained by the CARMENES\footnote{Calar Alto high-Resolution search for M dwarfs with Exoearths with Near-infrared and optical Échelle Spectrographs.} spectrograph \citep{bello-arufe_transmission_2023} and the combined spectrograph GIARPS \citep[GIANO-B + HARPS-N;][]{sicilia2024gapsprogrammetngxxx}. While the short transit baselines limited the capture of the outflow's full extent, both studies \citep{bello-arufe_transmission_2023, gully-santiago_large_2024} noted a helium redshift during the transit. Notably, Fig.~8 of \cite{bello-arufe_transmission_2023} shows greater pre-transit absorption depth, consistent with \cite{gully-santiago_large_2024}. 

%%% HAT-P-32b
The hot Jupiter HAT-P-32~b \citep{hartman_hat-p-32b_2011, wang_transiting_2019} orbits a late-F to early-G star ($T_{\rm eff} = 6001$~K) at a distance of $a = 0.03$~au, with an orbital period of $P_{\rm orb} = 2.15$ days and an equilibrium temperature of $T_{\rm eq} = 1835$~K. The transit duration is 3.12 hours, and the orbit has an eccentricity of $e \sim 0.16$.  The planet has a mass of $M_p = 0.68 M_J$ and a radius of $R_p = 1.98 R_J$, resulting in a surface gravity of $\log g_p = 2.6$~dex. 

The characteristics of HAT-P-32~b and HAT-P-67~b have many similarities. Due to their small orbital separation and the high mass ratio between the star and the planet, both planets are close to filling their Roche lobe, with $R_p / R_{RL} = 0.62$ for HAT-P-32~b and $R_p / R_{RL} = 0.37$ for HAT-P-67~b. CARMENES observations by \cite{czesla_h_2022} detected hydrogen (H$\alpha$) and helium (He-I) outflows from HAT-P-32~b. The spectral lines show irregular shapes and significant, time-dependent variations. 1D hydrodynamic modeling of these data by \cite{yan_possibly_2024} suggests that the planet's atmosphere has a solar metallicity. 1D Parker wind models by \cite{lampon_characterisation_2023} and 3D hydrodynamic simulations by \cite{zhang_giant_2023} estimate a mass-loss rate of approximately $10^{13}$\gs, similar to the estimate for HAT-P-67~b.

%% AIM
This paper aims to model  large-scale structures of asymmetric outflows as observed in HAT-P-67~b and HAT-P-32~b, with a particular focus on the conditions under which a high-density leading stream emerges. Our work tries to understand these systems' qualitative features rather than quantitatively fitting their observations. 

In Sect. \ref{sec:methods} we introduce the simulation setup and approach. Section \ref{sec:results} addresses the outflow conditions that lead to a leading tail only and explores the impact of the stellar wind on the outflow. Section \ref{sec:discus} places our results in context and discusses their broader implications. Finally, we present our conclusions in Sect. \ref{sec:concl}.

%--------------------------------------------------------------------
\section{Methods} \label{sec:methods}
% Big picture
To simulate the planetary outflow and its interaction with the stellar wind and the gravitational field of the host star, we used 3D hydrodynamical models and post-processed them to estimate the effects of the escaping planetary material on the transit spectra, focusing on the metastable helium triplet around $\lambda = 1083$~nm. The primary goal of this study is to identify the parameters responsible for the presence of a leading outflow stream, characterize the structural features of this planetary wind morphology, and determine how these features are reflected in helium transit observations. The simulations presented in this paper are modeled on parameters similar to those of HAT-P-67~b.

\subsection{Simulations}
% Overview / Code
We simulated a planet orbiting its host star, incorporating thermal winds from both the planet and the star using the 3D Eulerian (magneto)hydrodynamic code \texttt{Athena++}\footnote{\url{https://github.com/PrincetonUniversity/athena}}, version 2021 \citep{stone_athena_2020}. For these hydrodynamic simulations, we used the setup previously developed in \cite{macleod_stellar_2022} and extended to account for planetary day-night anisotropies in \cite{nail_effects_2024}. In the following, we outline the key features of our simulations and the modifications made for this study, building on the methods described in Sect. 2 of those works. 

% Assumptions
The simulation solves the equations for mass, momentum, and energy conservation in an inviscid gas. We used the ideal gas equation of state with an almost isothermal adiabatic index of $\gamma = 1.0001$, which results in nearly isothermal behavior along adiabats. However, temperatures at the points where the planetary winds originate from the surface can vary, allowing us to incorporate temperature differences between the planet's dayside and nightside, as described by \cite{nail_effects_2024}. We considered the gravitational influence of the planet and the star, while neglecting the back-reaction of the outflow on the planetary orbit and the impact of 1083~nm radiation pressure on the gas dynamics.

We inserted hydrodynamic winds into our simulations by parameterizing the conditions at the bases of the stellar and planetary winds. The pressure on the stellar surface is determined by the constant surface density, $\rho_*$, and the hydrodynamic escape parameter, $\lambda_*$, which represents the ratio of gravitational potential energy, $Gm M_*  R_*^{-1}$,  to thermal energy, $k_B T(R_*)$. Similarly, the pressure at the planetary surface is given by
\begin{equation}
    P_p = \rho_p \frac{GM_p}{\gamma \lambda_p R_p}~,
    \label{eq:press_p}
\end{equation}
where $\lambda_p = \frac{GM_p}{c_s^2 R_p}$ represents the hydrodynamic escape parameter for the planet. The surface densities of the planet ($\rho_p$) and the star ($\rho_*$) remain constant within each model and are provided in Table \ref{tab:Model_parameter}. In models that account for day-night anisotropy, the surface pressure can vary across the planetary surface. 

% Mesh / Parameter Choices
Using a spherical polar mesh with the star at the center of the corotating frame aligned with the planet's orbit, our computational domain extends from the star's surface to a distance of 12.3\,au in all directions. The planet is located on the negative x-axis at a distance of $a = 0.062$\,au from the star. Both objects rotate with the orbital frequency of the planet, $\Omega = \Omega_{\text{orb}} = G(M_p+M_*)a^{-3}$, with their angular momenta aligned in the positive z direction. The base mesh is composed of $12 \times 8 \times 12$ mesh blocks, each containing $16$ zones, logarithmically spaced in the $r$ direction and evenly spaced in $\theta$ and $\varphi$ directions, maintaining nearly cubic zone shapes. We reduced the effective zones in the $\varphi$ direction near the poles by averaging conserved quantities to avoid extreme aspect ratios.

To assess the sensitivity of our results to numerical parameters, we varied the spatial resolution with three refinement levels around the planet (increasing by a factor of two) and three levels in the outflow torus. We find that stellar and planetary outflow rates stabilize after a few planetary orbits and converge above the lowest resolution levels ($N_{\mathrm{SMR}} = 3$). However, spectral properties, particularly in the extended tails, require higher resolution. Thus, we used $N_{\mathrm{SMR}} = 5$ for the region around the planet and $N_{\mathrm{SMR}} = 2$ for the tail in our fiducial settings.

We investigated six models with varying planetary outflow temperatures and geometries, as well as stellar mass-loss rates. Detailed input parameters are provided in Table \ref{tab:Model_parameter}. Models designated with a prefix "B" have a low value of $\lambda_p =3$, \citep[assuming a gas temperature of 14000~K (based on 1D model results from ][]{gully-santiago_large_2024} and a mean molecular weight of 0.6, consistent with system properties and uncertainties. However, our simulations indicate that helium line broadening can be kinematic rather than purely thermal, motivating the exploration of cooler winds. To test this, we introduced models, prefixed with "S," that assume a colder planetary wind with $\lambda_p = 9$, corresponding to a temperature of approximately 5000 K, though it could theoretically be even lower.

Models B1 and S1 assume an isotropic planetary outflow, whereas all other models incorporate a dayside dominant outflow configuration using the anisotropic planetary wind model introduced in our previous work \cite{nail_effects_2024}, detailed in Sect. 2.2. In these anisotropic scenarios, the pressure on the nightside of the planet is only 1\% of that on the dayside, as described by the parameter $f_{\rm press}$. Models S3 and S4 are similar to the anisotropic cold model S2, but we gradually increased the stellar mass-loss rate. The system parameters, adopted from HAT-P-67~b, are identical for all models: 
$a = 0.062$~au = $9.275 \times 10^{11}$~cm, 
$M_* = 1.37~M_{\astrosun} = 2.733\times 10^{33}$~g, 
$R_* = 2.64~R_{\astrosun} = 1.84\times 10^{11}$~cm, 
$M_p = 0.34~M_{\rm J}= 6.45 \times 10^{29}$~g, and 
$R_p = 2.15~R_{\rm J} =  1.54 \times 10^{10}$~cm.

\begin{table*}
    \caption{Input parameters for each simulation model.}
     \centering
     
    \begin{tabular}{lccccccc}
    \toprule
    \toprule
         model & $\lambda_p$ & $\rho_p$ & $\rho_*$ & $f_{\rm press}$ & $T_{\rm sub}$ & $c_{\rm s, sub}$  & $c_{\rm s, anti}$ \\ 
          & & [g~cm$^{-3}$] & [g~cm$^{-3}$] & [\%] &  [K] & [km~s$^{-1}$] & [km~s$^{-1}$] \\  
    \midrule
         B1 & 3 & $6.92 \times 10^{-15}$ & $4.64 \times 10^{-15}$ & 100 & 14\,100 & 9.7 & 9.7 \\
         B2 & 3 & $6.92 \times 10^{-15}$ & $4.64 \times 10^{-15}$ & 1   & 14\,100 & 9.7  & 1.0 \\
         \midrule
         S1 & 9 & $5.38 \times 10^{-13}$ & $4.64 \times 10^{-15}$ & 100 & 4\,700 & 5.6 & 5.6 \\
         S2 & 9 & $5.38 \times 10^{-13}$ & $4.64 \times 10^{-15}$ & 1 & 4\,700 & 5.6  & 0.6 \\
         \midrule
         S3 & 9 & $5.38 \times 10^{-13}$ & $2.09  \times 10^{-14}$ & 1 & 4\,700 & 5.6  & 0.6 \\
         S4 & 9 & $5.38 \times 10^{-13}$ & $2.78 \times 10^{-14}$ & 1 & 4\,700 & 5.6  & 0.6 \\
    \bottomrule
        \end{tabular}
        
\vspace{1mm}
\justifying \noindent \small Notes. We present two hot wind models (B) and four cold wind models (S). For each, $\lambda_p$ is the hydrodynamic escape parameter, and $f_{\rm press}$ is the pressure fraction at the anti-stellar point relative to the substellar point. The table lists the planetary and stellar surface density ($\rho_p$ and $\rho_*$, respectively); the substellar temperature $T_{\rm sub}$, calculated with $\mu = 1.25$ (although this varies as a function of the outflow's location); and the sound speed $c_s$ at both substellar (sub) and anti-stellar (anti) points on the planetary surface, which has a constant density $\rho_p$. Models S3 and S4 have a higher stellar mass-loss rate compared to model S2. 
\label{tab:Model_parameter}
\end{table*}

\subsection{Radiative transfer}

In our radiative transfer analysis, we adopted the approach outlined by \citet{macleod_stellar_2022}. We assumed a solar gas composition with mass fractions of hydrogen, helium, and metals represented by X = 0.738, Y = 0.248, and Z = 0.014, respectively, throughout the computational domain. We assumed a detailed equilibrium in each simulation grid cell, treating the system as being in a steady state. To compute the unattenuated photoionization rate, $\Phi$, we adopted an intermediate flux between the low and high extreme-ultraviolet (XUV) synthetic spectra of HAT-P-67 as presented in \citet[see their Fig.~12]{gully-santiago_large_2024}. Combined with the optical depth $\tau$, this yields the photoionization rate $\Phi e^{-\tau}$. The optical depth to hydrogen-ionizing photons is iteratively calculated for each cell in the simulation grid.

To generate synthetic helium spectra, we computed the integrated optical depth along rays extending from the stellar surface to the observer in a simulation snapshot. A Voigt line profile is imposed, incorporating a Gaussian component that is dependent on the local gas temperature. Rays crossing the planet's interior are excluded, and the contribution of rays is weighted according to the stellar quadratic limb darkening law. An impact parameter of 0.5, as determined by \cite{gully-santiago_large_2024}, is used in the calculations of the spectra. Rotational effects of the stellar disk are not considered.

Due to the larger optical depths in our simulations, more iterations are required to determine the ion populations compared to \citet[see their Eqs. 9 and 10]{macleod_stellar_2022}. The number of required iterations increases with the up-scaling of the density in the snapshot. Therefore, we used an iterative scheme to ensure convergence of level populations even when the optical depth is of order unity. The code updates the densities at each iteration based on radiative transfer calculations, including photoionization and recombination processes. The solution is updated until the convergence criterion is met, specifically when the relative change in the ionized hydrogen density, becomes smaller than a predefined threshold (here 0.02\%). 

%--------------------------------------------------------------------
\section{Results} \label{sec:results}
\subsection{Formation of a predominantly leading stream}

\begin{figure}
    \centering
    \includegraphics[width=0.49\textwidth]{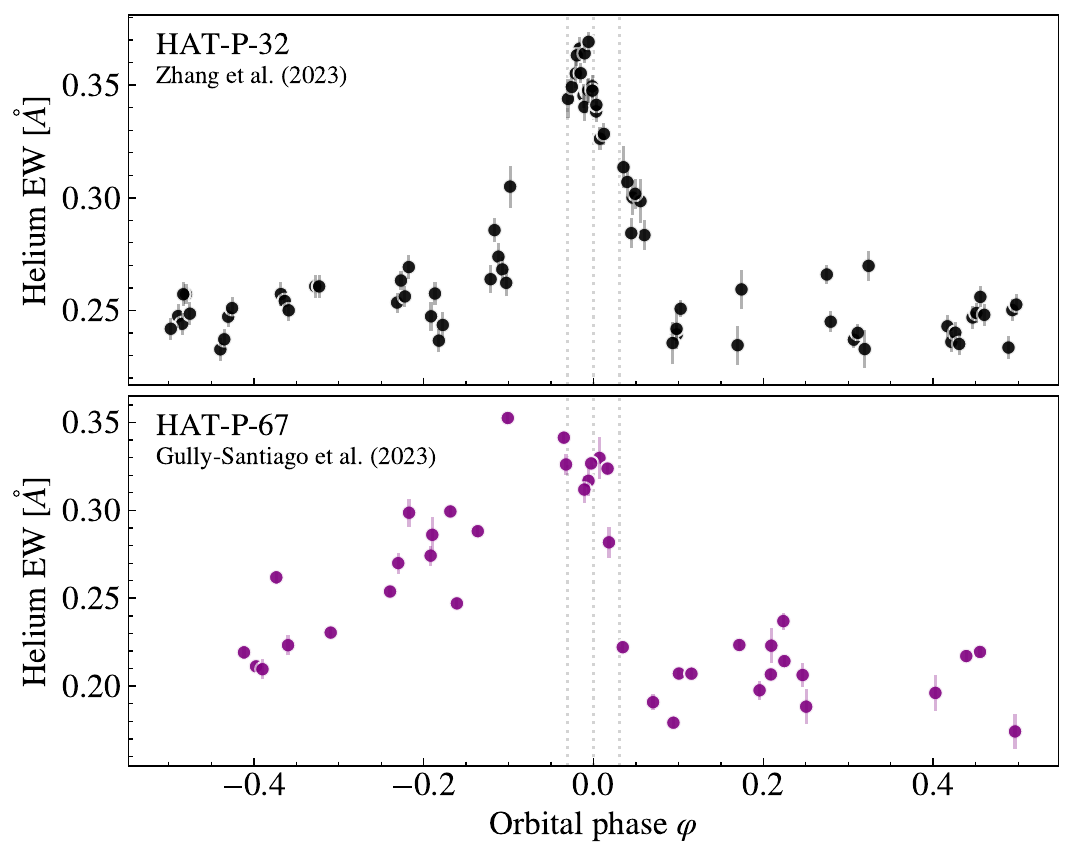}
    \includegraphics[width=0.49\textwidth]{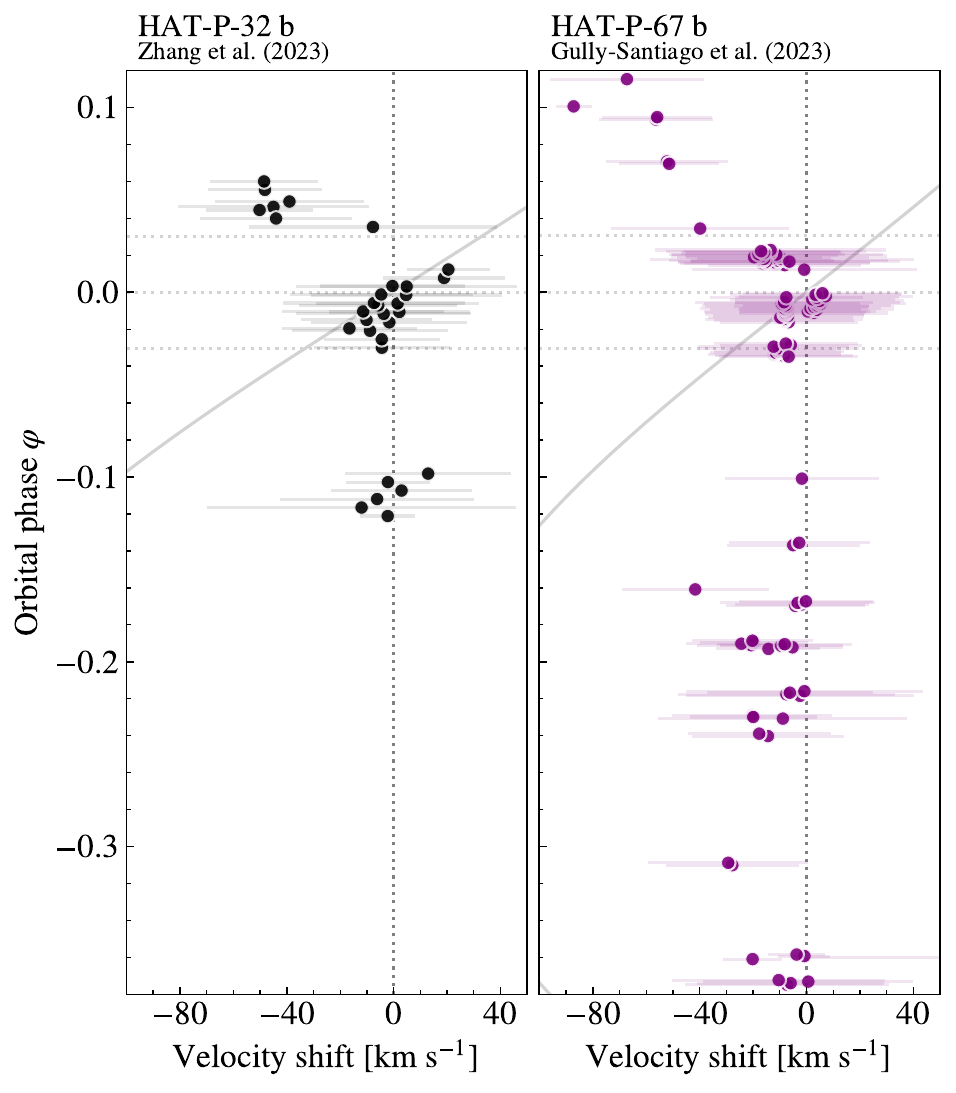}
    \caption{Observational results from the HPF transit observations across the full orbit of HAT-P-32~b \citep{zhang_giant_2023} and HAT-P-67~b \citep{gully-santiago_large_2024}. Top two panels: \hei EWs as a function of orbital phase. The dashed gray lines mark the beginning, middle, and end of the optical transits. Bottom two panels: Velocity shifts of the planetary \hei excess absorption only across the orbital phase. The solid gray line represents the planetary orbital motion, as the velocity shifts are measured in the stellar rest frame. Both planets show a longer pre-transit absorption than  post-transit absorption, and the \hei excess absorption is highly blueshifted for the pre-transit phases.}
    \label{fig:HPF_obs}
\end{figure}

\begin{figure*}
    \centering
    \includegraphics[width=.77\textwidth]{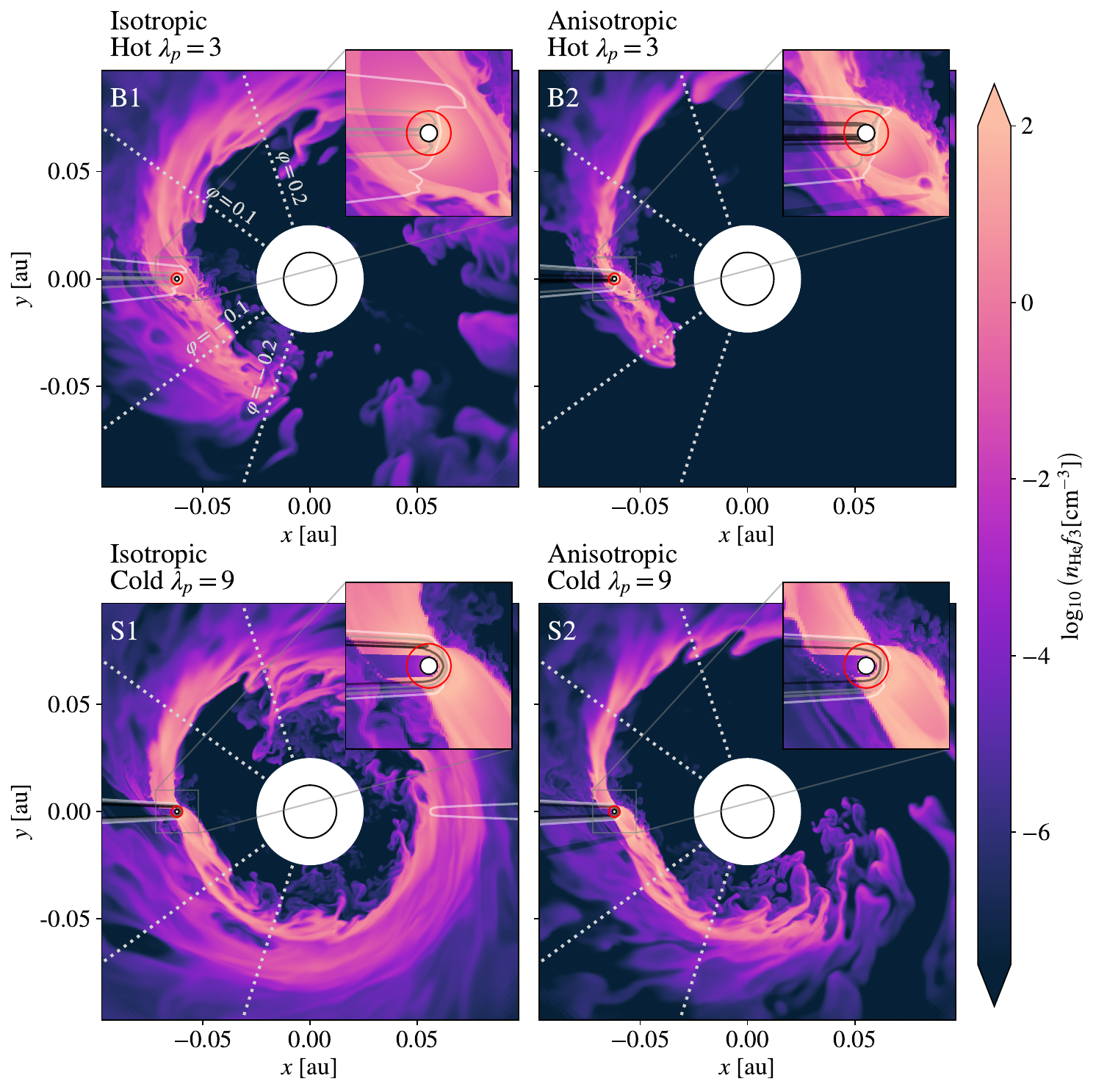}
    \caption{Metastable helium number density derived from the planetary wind, contrasting hot and cold planetary winds alongside isotropic and dayside-dominant outflow configurations. The star is positioned at the center, with the planet located in the $-x$ direction, orbiting the star counterclockwise. The black circles indicate their extent, while the red circle represents the Hill radius, $R_H = 2.6~R_p$, of the planet. Black, gray, and white contours mark the cumulative optical depth of $\tau = 10, 1$, and $0.1$, respectively, of the metastable helium line toward the observer in the radial direction. The dotted lines mark the orbital phase angles, $\varphi$. All snapshots are taken after eight orbits, with the steady state reached after four orbits. The time evolution of model S2 is illustrated in the  movie provided in Appendix \ref{sec:appA}. Only the scenarios involving a cold planetary wind result in a widely extended, high-density outflow. Specifically, a cold, dayside-dominated outflow (S1) produces a more prominent leading stream compared to the trailing one.}
    \label{fig:IsoAniso}
\end{figure*}

\begin{figure}
    \centering
    \includegraphics[width=.49\textwidth]{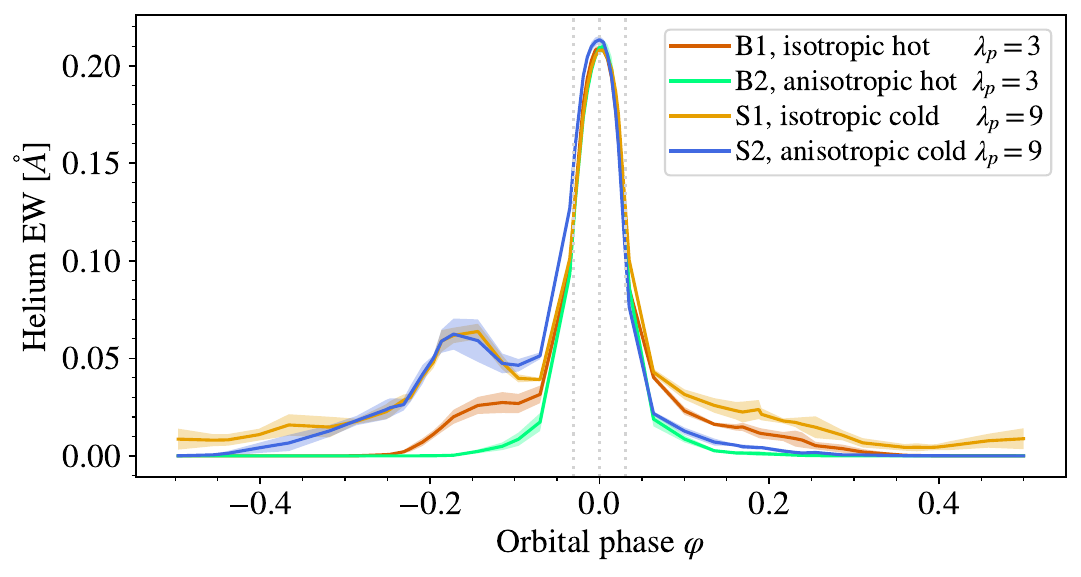}
    \caption{EW of the \hei line across orbital phases, comparing model results from Fig.~\ref{fig:IsoAniso}. Shaded regions indicate the standard deviation calculated from five spectra per orbital phase. Horizontal lines denote the first and fourth contact of the transit. A cold, dayside-dominant outflow from the planet is essential to reproduce an asymmetric light curve as observed for HAT-P-67~b and HAT-P-32~b}
    \label{fig:EW}
\end{figure}

\begin{figure*}
    \centering
    \includegraphics[width=.87\textwidth]{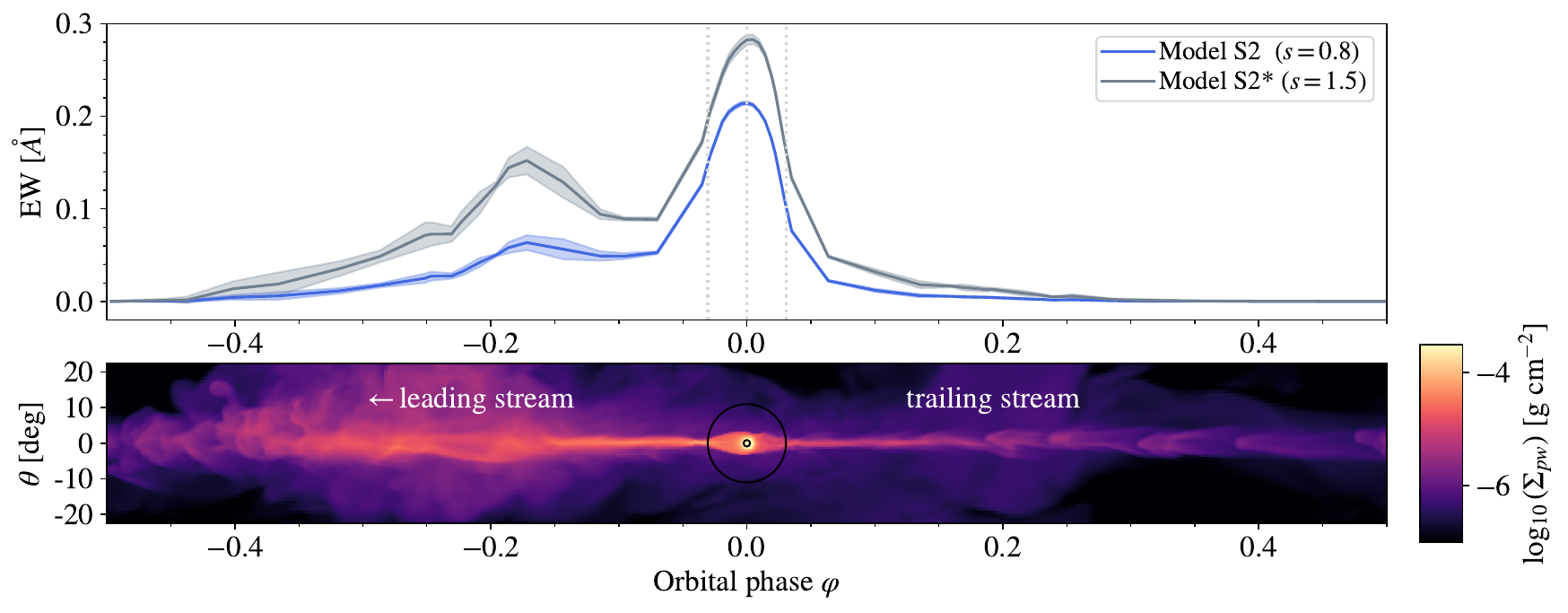}
    \caption{EW of the helium 1083~nm line across orbital phases (top) in comparison to the surface density projected along the radial direction of the planetary wind, $\sum_{pw}$, along the orbit of model S1 (bottom). In the spectral light curve, we show two different density scalings, $s$, of the same simulation snapshot: one that matches the in-transit absorption (S2, blue) and one that fits the post-transit absorption (S2*, gray). The black circles indicate the angular sizes of the planet and the star. The local maximum at $\varphi = -0.2$ in the light curve can be explained by the large vertical extent of the gas, a result of instabilities induced by the stellar wind.}
    \label{fig:surface_density}
\end{figure*}

In this section we systematically explore a range of parameters in our simulations to generate a high-density leading outflow stream. Our analysis is based on a comparison with the \hei observations of HAT-P-67~b and HAT-P-32~b, shown in Fig.~\ref{fig:HPF_obs}. Typically, planetary excess absorption is isolated by calculating the difference between the observed absorption during transit and the baseline stellar absorption observed out of transit. The residual line refers to the observed absorption that remains after accounting for the baseline. This method separates the contribution from the planet's atmosphere, enabling the measurement of absorption by different atmospheric species. However, if the baseline observed out of transit is contaminated, it could lead to a misinterpretation of the planetary absorption. Therefore, the top panels of Fig.~\ref{fig:HPF_obs} present the combined \hei absorption equivalent width (EW) from both the star and the planet. These plots demonstrate that the planetary excess absorption persists even after the transit.

Notably, the pre-transit absorption, likely caused by a high-density leading stream, persist longer than the post-transit absorption in both planets. In the bottom panel, we present the observed velocity shifts of the residual planetary helium line. The velocity shift of the residual helium line refers to the displacement of the observed absorption feature from the expected rest wavelength of the helium line, which indicates the motion of the planetary atmosphere. A blueshift means that the gas is moving toward the observer, while a redshift indicates that the gas is moving away. The residual helium line, of HAT-P-67~b and HAT-P-32~b, is significantly blueshifted during pre-transit observations. While the in-transit observations seem to be variable and partly follow the planetary movement, the post-transit spectra do not. Instead, they either remain in the stellar rest frame or shift slightly toward shorter wavelengths in the later phases. 

To understand the factors influencing the formation of an extended leading outflow, we compared and analyzed models B1, B2, S1, and S2. In Fig.~\ref{fig:IsoAniso} we present their outflow morphologies, illustrating the metastable helium number density of the planetary wind in the orbital midplane as calculated from the radiative transfer analysis. In Fig.~\ref{fig:EW} we display the corresponding light curves, showing the EW of the helium line as a function of orbital phase. To calculate the EW, we integrated the total flux of the helium triplet over the wavelength range of 10827.00 to 10831.75 $\AA$. Unlike the light curves shown at the top of Fig.~\ref{fig:HPF_obs}, here we display only the planet's excess absorption, with the stellar contribution excluded.

Due to the scale-invariance of hydrodynamic equations, we can uniformly adjust the pressure and density of a simulation snapshot during post-processing without altering the outflow morphologies. This allows us to modify the planetary and stellar mass-loss rates, without changing their ratio, before generating the synthetic spectra. We used this capability to match the model light curve to the mid-transit absorption of HAT-P-67~b. We provide the resulting mass-loss rates of these models in Table \ref{tab:alter_model_param}. Mass-loss rates are determined by tracking the mass changes of the planet and star over time and accounting for mass flux through the simulation boundaries. This method provides a net mass-loss rate for each object, ensuring mass conservation while accounting for mass ejected, accreted, or redistributed within the system. We focused on the steady-state regime by averaging the rates over times greater than $10^6$~s, scaling the results by the predefined scaling factor $s$ for each simulation.

\begin{table}[h!]
    \centering
    \caption{Time-averaged planetary and stellar mass-loss rates from simulations. }
    \begin{tabular}{lccc}
        \toprule
        \toprule
         model  & $s$ & $\langle \dot{m}_p \rangle $ &  $\langle \dot{m}_* \rangle$ \\ 
          & & [g~s$^{-1}$] & [g~s$^{-1}$]\\ 
         \midrule
         B1  & 1.40 & $1.1 \times 10^{13}$ &  $1.2 \times 10^{14}$\\
         B2  & 2.50 & $5.2 \times 10^{12}$ & $2.3 \times 10^{14}$\\
         \midrule
         S1 & 0.36 &  $5.2 \times 10^{12}$ &  $2.7 \times \times 10^{13}$\\
         S2 & 0.80 & $5.1 \times 10^{12}$ & $6.7 \times 10^{13}$\\
         S2* & 1.50 & $9.6 \times 10^{12}$ & $1.3 \times 10^{14}$\\
         \midrule 
         S3 & 4.10 & $2.1 \times 10^{13}$ & $1.7 \times 10^{15}$\\
         S4 & 5.50 & $2.6 \times 10^{13}$ & $3.1 \times 10^{15}$\\
         \bottomrule
    \end{tabular}
    
    \justifying Notes. The values are measured when the quasi-steady state in the simulation is reached, $t > 1.66$~s. These rates incorporate the density scaling $s$ necessary to match the mid-transit absorption of HAT-P-67~b. 
    
    \label{tab:alter_model_param}
\end{table}

Firstly, we examined model B1, which features a hot planetary isotropic outflow (Fig.~\ref{fig:IsoAniso}, top left). In this configuration, the strong planetary wind creates a cavity against the stellar wind pressure. Due to the high temperature of the outflow, $T(r_H) \approx 7000$~K at the Hill radius $r_H$ of the planet, the radial velocity component $v_r(r_H) = 16$~\kms dominates over the shear velocity $v_{\varphi}(r_H) = 1$~\kms, resulting in a bubble-shaped morphology. Optical depth contours of metastable helium reveal a symmetric density distribution around the $y = 0$ axis and the shadow of the planet, in the immediate vicinity of the planet, inside the unshocked cavity. Beyond the cavity, both leading and trailing outflows emerge. However, the density in the extended outflow is much lower than in the vicinity of the planet. This is reflected in the light curve, shown by the red line in Fig.~\ref{fig:EW}. 

In an attempt to produce an asymmetric light curve with higher pre-transit absorption, we implemented a dayside-dominated outflow compared to the first model. This second model, B2 (Fig.~\ref{fig:IsoAniso}, top right), is similar to model B1 but features a more extended cavity on the dayside compared to the nightside. As discussed in \cite{nail_effects_2024}, a net day-to-nightside flow arises due to the pressure gradient from the substellar to anti-stellar point, resulting in increased density on the trailing side. Additionally, the pressure in the leading arm is insufficient to counteract the stellar wind. Consequently, the gas is pushed to the trailing side of the planet, contributing to the extended trailing arm. However, the density in the trailing side is too low to contribute significantly to post-transit observations (Fig.~\ref{fig:EW}, green line).

A distinct morphology emerges in models S1 and S2, featuring a colder planetary wind, corresponding to a high hydrodynamic escape parameter ($\lambda_p = 9$). Unlike the morphologies previously described, high-density streams surround the planet (Fig.~\ref{fig:IsoAniso}, bottom). The densities in the immediate vicinity of the planet are so high that a portion of the XUV radiation is blocked, resulting in a low ionization fraction and, consequently, a low population of metastable helium. The dayside outflow launched toward the star, with $v_r(r_H) = 7$~\kms and $T(r_H)\approx 4300$~K, crosses the inner Lagrange $L_1$ point, and is drawn into a Keplerian orbit after overcoming the gravitational potential of the planet. The outflow can potentially extend over hundreds of planetary radii along the planet's orbit. In the case of a hot wind, the radial velocity component is sufficiently strong to prevent the gas from being pulled back into orbit after passing the $L_1$ point. On the nightside of the planet In the cold wind scenario, the outflow accelerates in the negative x direction, passes through the $L_2$ Lagrange point, and undergoes similar entrainment into a Keplerian orbit. Since the $L_2$ point is further outward relative to the orbit than $L_1$, the trailing gas moves more slowly, while the gas leading the planet moves faster. 

While the isotropic model S1 produces a more symmetric outflow surrounding the planet, resulting in a toroidal structure around the star, model S2 demonstrates a higher density in the leading stream. Model S2 is the only configuration that displays significantly greater pre-transit absorption compared to post-transit absorption (see the blue light curve in Fig.~\ref{fig:EW}). Notably, the helium EW ratio between orbital phases $-0.2$ and $0.2$ is highest for model S2, with a value of 9.8, while model S1 shows a significantly lower ratio of 1.9. When comparing this to the observational results for HAT-P-67~b (see Fig.\ref{fig:HPF_obs}), and assuming, with some uncertainty, that the planet’s excess absorption can be approximated by subtracting a baseline value of 0.2~$\AA$ from the combined stellar and planetary helium EW light curve, the resulting ratio is approximately 7, which is closer to the value predicted by model S2. In conclusion, our findings suggest that a leading tail extending over hundreds of planetary radii is achieved when a cold planetary wind originates exclusively from the dayside. Simulation model S2 seems to best reproduce this behavior; however, the pre-transit absorption seems to be underestimated compared to the observational results shown in the top panel of Fig.~\ref{fig:HPF_obs}. 

In addition to investigating the orbital midplane, we also examined the vertical structure of the planetary wind. Figure~\ref{fig:surface_density} presents the helium EW light curve along with the surface density of the planetary wind throughout the planetary orbit of the fiducial model S2. We show one light curve resulting from a density scaling factor of $s = 0.8$ and also include results from a higher scaling factor of $s = 1.5$. While the latter overestimates the in-transit absorption, it more accurately matches the tail absorption. A possible explanation for why our models cannot fully capture the density gradient between the tails and the planet is that they do not account for thermodynamics, cooling or heating rates.

Looking at the vertical structure, the outflow from the dayside of the planet converges at the phase $\varphi \sim -0.05$ due to tidal compression, covering a smaller part of the stellar disk and causing a dip in the light curve. As the phase progresses earlier, the outflow disperses into irregular patches and covers a larger part of the stellar disk, which explains the local maximum in the light curve at $\varphi \sim -0.19$. The density distribution is highly variable due to instabilities, which is reflected in both the observed light curve and our model. In our model, the orbit-to-orbit variability is about 20 m$\AA$ in EW, while the observed spread can reach up to $\sim 100$ m$\AA$ (e.g., at $\varphi \sim -0.22$ for HAT-P-67~b).

%--------------------------------------------------------------------
\subsection{Characteristics of the helium line profile caused by the stream morphology}

\begin{figure}
    \centering
    \includegraphics[width=.42\textwidth]{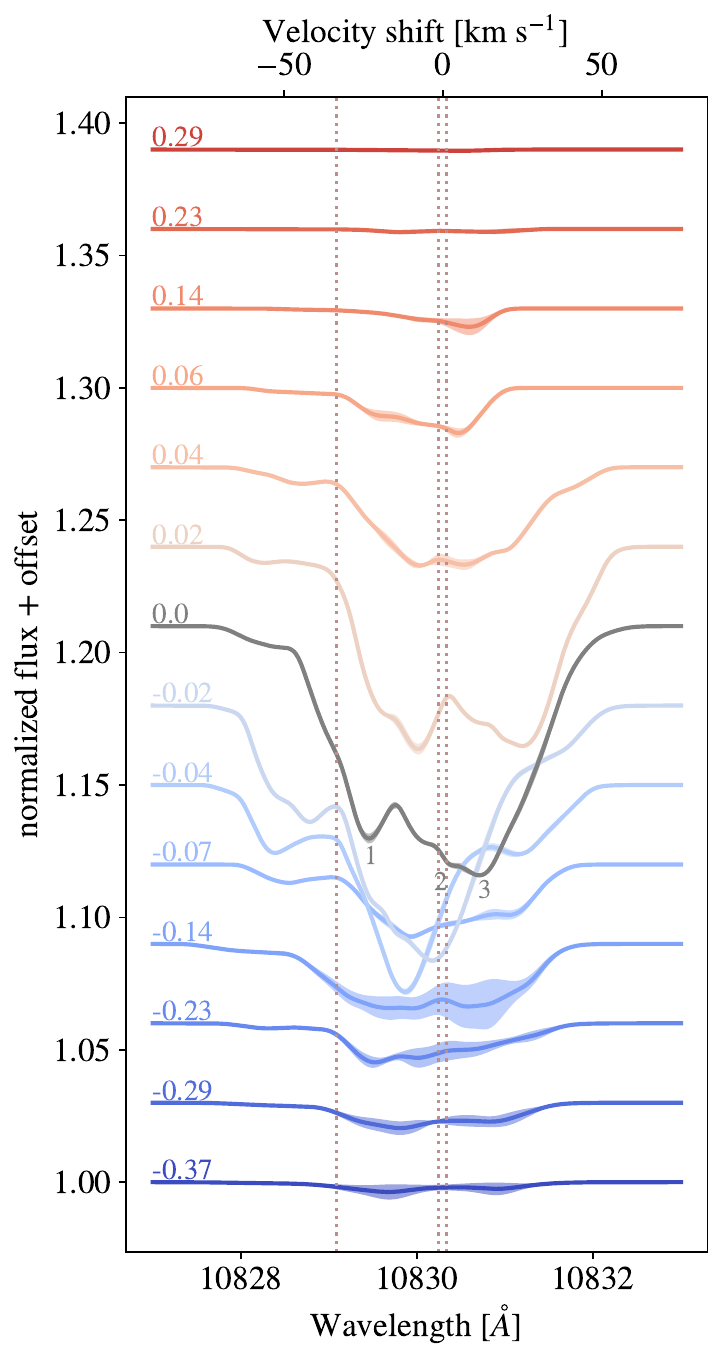}
    \caption{Spectral time series showing the helium triplet (model S2) in the stellar rest frame. The orbital phase is denoted on the top left of each spectrum, while the transit time progresses from bottom to top. Dotted lines highlight the nominal wavelengths of the helium triplet in air. The solid lines show the average computed over five spectra from snapshots taken after one orbit; shaded regions represent the standard deviation. The pre-transit spectra (blue) show significant time variability. In-transit spectra (phases -0.03 to 0.03) show a highly irregular shape as they are kinematically and not thermally broadened. Three characteristic peaks are indicated with numbers. The post-transit absorption (red) is minimal or absent for larger phases.}
    \label{fig:spectra}
\end{figure}

In Fig.~\ref{fig:spectra} we present the spectra of the fiducial model S2 in sequential order, progressing from pre-transit (blue) to mid-transit (gray) and post-transit (red). The line profiles show pronounced irregularities, deviating significantly from a thermally broadened Gaussian profile. The mid-transit line profile reveals three distinct features, each marked with corresponding numbers. Based on the velocity distribution observed in the snapshots of the stream morphology (see Fig.~\ref{fig:v-map}), we expect the \hei feature to show kinematic broadening, with three distinct peaks. These peaks would correspond to different velocity regimes within the planetary wind, in addition to the \hei feature being a triplet. The first, blueshifted peak is formed by gas launched from the nightside that moves toward the observer. The third peak is redshifted, originating from the dayside wind moving away from the observer. Additionally, there is a line component with little to no shift, which forms in a high-density region close to the planet's surface, resulting in the second peak.

This is what we observe in Fig.~\ref{fig:spectra}; however, the formation of the first peak is actually more complex. It is not just a part of the helium main component (the unresolved lines of $\lambda_{\rm air} = 10830.25\AA$ and $10830.34~\AA$) that is blueshifted as is forms in the gas launched from the nightside; the blue component ($\lambda_{\rm air} = 10829.09~\AA$) of the third peak also contributes to the first peak. As the helium line forming in a high-density region is optically thick, the blue component is significantly enhanced. Consequently, the first peak is a combination of the blueshifted main component and the redshifted blue component from the third peak. In Appendix \ref{sec:line_formation} we analyze the line-forming regions, focusing exclusively on the singlet line to isolate the contributions of the triplet components and the three kinematic peaks.

The orbit-to-orbit variability, as indicated by the standard deviation calculated from the average of five snapshots, is low, while the three aforementioned features persist. It is also notable that the third peak of the mid-transit spectrum shows the highest absorption. This is due to the day nightside anisotropy, with higher gas density on the dayside of the planet. If features 2 and 3 are not observationally resolved, the overall line may appear redshifted, while the exact position and strength of the three peaks depend on the density and velocity distribution of the specific planetary wind. Notably, the shape of the highly resolved \hei line observed in HAT-P-32~b by \citet[see their Fig.~18]{czesla_h_2022} shows similar irregularities, indicating a kinematically broadened line. While instrumental uncertainties cannot be entirely ruled out, the blue component of the triplet appears to be double-peaked, and on the longer wavelength side of the main component, two smaller peaks are visible. The small ratio between the blue and main components further suggests that the line is optically thick.

For our fiducial model S2, the absorption after the transit (red spectra) is only significant up to the phase $\varphi\sim 0.05$. At higher orbital phases, the absorption diminishes, aligning with the sharp edge in the light curve (see Fig.~\ref{fig:EW}). In the snapshot of simulation S2 (Fig.~\ref{fig:IsoAniso}, bottom right) a trailing tail is visible; however, the gas density probed at higher phase angles is not high enough to contribute to the line formation. As illustrated in Fig.~\ref{fig:EW}, pre-transit absorption extends up to $-0.3$ in orbital phase. In contrast to the in-transit spectra, during the pre-transit phase ($\varphi < -0.03$), both the line profile and EW display high variability. This indicates that the leading arm is exposed to instabilities from the interaction with the stellar wind. The snapshot in Fig.~\ref{fig:IsoAniso} indeed shows Kelvin-Helmholtz and Rayleigh-Taylor instabilities at phases $\varphi < -0.1$. 

Furthermore, we explored the line profiles for model S1, the isotropic stream scenario. In this case, the three peaks during transit are also visible, with the key difference being that the post-transit spectra show higher absorption depths. However, the overall shapes of the lines remain unchanged. In the isotropic case (S1), the overall spectrum appears less redshifted compared to the anisotropic case (S2). This is because the cells contributing to line formation are more symmetrically distributed around the planet in model S1. In contrast, in the anisotropic case (S2), the cells are predominantly located in the leading stream, where higher densities occur due to the day-night anisotropy (see Fig.~\ref{fig:v-map}).

%--------------------------------------------------------------------
\subsection{Wind--wind interactions' influence on tails}

\begin{figure*}
    \centering
    \includegraphics[width=.85\textwidth]{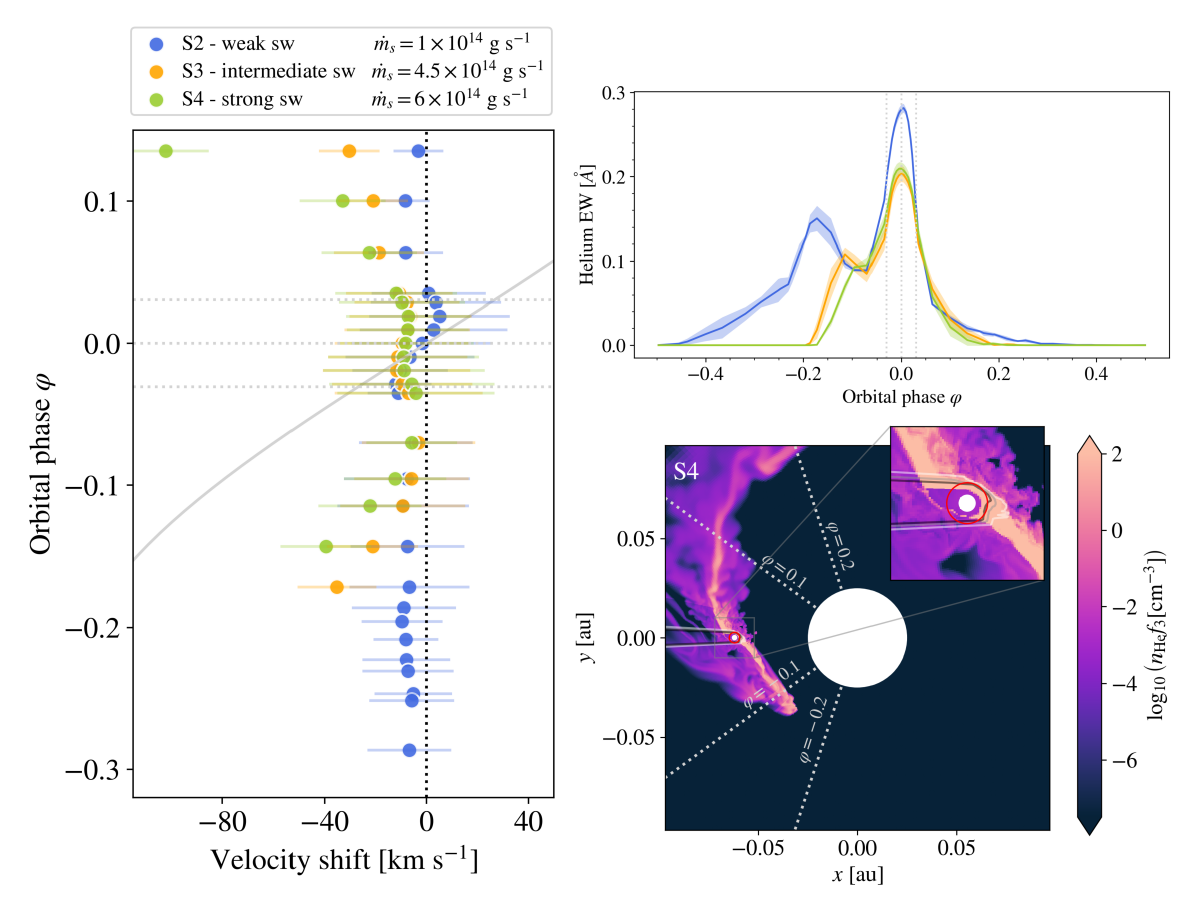}
    \caption{Impact of stellar wind (sw) on planetary outflow and \hei line velocity shifts. Left: Helium line velocity shift in the stellar rest frame across orbital phases for models S2, S3, and S4. The error bars represent the full width at half maximum, adjusted by dividing by $2\sqrt{2\ln{2}}$ to approximate the standard deviation of a Gaussian distribution. The vertical gray line indicates the planetary movement. Top right: Corresponding light curves. Bottom right: Metastable helium number density in the orbital midplane under a strong stellar wind influence (model S4). Black, gray, and white contours mark the cumulative optical depth of $\tau = 10, 1$, and $0.1$, respectively, of the metastable helium line toward the observer in the radial direction. The helium blueshift increases with increasing stellar mass-loss rate, providing insights into wind kinematics. A strong stellar wind diminishes the planetary outflow, making the light curve a more robust analysis tool for characterizing the atmospheric escape. }
    \label{fig:stellar_wind+shift}
\end{figure*}

Similar to the studies by \cite{mccann_morphology_2019} and \cite{macleod_stellar_2022}, we investigated how varying stellar wind strengths (weak, intermediate, and strong) affect planetary wind morphology. To this end, we tested two additional models, S3 and S4, which share the same parameters as our fiducial model S2 but with progressively increased stellar mass-loss rates. Our results reflect those of the aforementioned studies: a weak stellar wind allows the planetary outflow to extend through the orbit, forming a torus around the star. If the stellar wind is sufficiently strong, the planetary wind becomes confined, forming a trailing tail behind the planet \citep[see also][]{ehrenreich_giant_2015, carroll-nellenback_hot_2017, vidotto_stellar_2020, carolan_effects_2021}. 

Furthermore, we find that the velocity shift of the metastable helium line provides valuable insights into constraining the strength of the stellar wind. The left panel in Fig.~\ref{fig:stellar_wind+shift} illustrates the velocity shift of the helium line across different stellar wind scenarios throughout the orbital phase. Given the highly irregular line profiles, we opted to calculate the weighted mean of the spectra to determine the velocity shift. Each wavelength is assigned a weight based on its corresponding absorption depth. Using this weighting, a weighted average wavelength is calculated, representing the central position of the helium spectrum. This wavelength, representing the central position of the helium spectrum, is then converted into a velocity shift relative to the centroid of the helium triplet's main component.

In the case of a weak stellar wind (model S2, blue), the velocity shifts during mid-transit follow the orbital movement of the planet. During ingress, the helium line is redshifted relative to the planetary movement, while during egress it is blueshifted. These shifts suggest that, during ingress, the probed gas in the leading stream is accelerated away from the planet toward the star (toward the $L_1$ point). Conversely, during egress, the gas in the trailing stream is accelerated away from the planet toward the observer (toward the $L_2$ point). 

After passing through the Lagrangian points, the gas forms tidal streams that orbit the host star in a Keplerian manner, with low radial velocities compared to the planet’s velocity. The absorption signatures of these streams therefore remain close to the stellar rest frame. In theory, the gas in the leading stream moves with a higher angular velocity as it orbits the star at a shorter distance. Similarly, the gas in the trailing stream moves slightly more slowly, orbiting at a greater distance. 

With increasing strength of the stellar wind, both pre- and post-transit spectra show progressively greater blueshifts. In contrast to S2, the velocity shifts in the transit of the stronger stellar wind models S3 and S4 no longer follow the planetary motion. Due to the high ram pressure of the stellar wind, the planetary outflow is strongly confined. As a result, the metastable helium densities around the planet become so high that the line primarily forms in the turbulent region farther from the planet, closer to the star (see the optical depth contours in Fig. \ref{fig:stellar_wind+shift}, bottom right). This gas is pushed toward the observer by the stellar wind, similar to the tails, resulting in a blueshift of the spectral line.

Furthermore, if the stellar wind is too strong, it tightly confines the planet's outflow, as shown in the bottom right panel of Fig.~\ref{fig:stellar_wind+shift}. This suppression notably affects the leading arm, resulting in shortened pre-transit absorption. Therefore, based on the light curves, we consider model S2 as the most suitable scenario explored in this work for explaining the observations of HAT-P-67~b and HAT-P-32~b (see Fig.~\ref{fig:HPF_obs}). However, looking at the velocity shifts, model S2 cannot reproduce the high blueshifts of $\sim -40$\kms\ for orbital phases $\varphi > 0.04$ while the stronger wind scenarios can. 

This analysis emphasizes the significance of two key measurements to constrain the planetary outflow with our simulations. Both the EW and the velocity shift over orbital phase are important. The light curve reveals the extent of the outflow, while the velocity shift provides insights into its kinematics. Importantly, the velocity shift is intricately linked, as demonstrated, to the strength of the stellar wind. However, the light curve serves to rule out a stellar wind scenario that is too strong, thereby establishing it as the more robust analysis tool.

\section{Discussion} \label{sec:discus}

% what are differences to Nail 2024?
In comparison to our previous work \citep{nail_effects_2024}, which explored the effects of anisotropic outflows in a hot wind scenario, this study focuses on the behavior of a cold planetary wind. In \cite{nail_effects_2024}, a significantly enhanced dayside outflow resulted in a more extended cavity around the planet before the gas was shocked and pushed behind the planet by the stellar wind. Additionally, a correlation was observed between the blueshift of the metastable helium line and the degree of anisotropy. In contrast, for the cold planetary wind studied here, the spectral lines are broadened kinematically. During transit, this broadening is primarily driven by the two streams diverging in opposite directions, whereas outside of transit, the line shift is mainly influenced by the strength of the stellar wind.

% why is the outflow cold? 
The question arises as to why the planetary winds of HAT-P-67~b and HAT-P-32~b should be cold, with temperatures of only a few thousand kelvins, rather than hot, reaching around 10000~K A contributing factor for a cooler outflow is that the effective potential for overcoming the Roche lobe is relatively low, since the planets are close to their host stars while having a low gravitational potential. The gas can escape more easily from the Roche lobe and is not further heated due to the expansion cooling. 

% why is the outflow asymmetric?
The second question that arises is why the outflow is asymmetric, with a more pronounced leading stream. While we remain agnostic about the wind-driving mechanisms in this study, it becomes clear that a dayside-dominated outflow can reproduce an asymmetric light curve. This suggests that stellar radiation indeed plays a role in driving the escape. Further supporting this theory is the observation that the outflow of HAT-P-67~b has apparently a greater degree of asymmetry than HAT-P-32~b. HAT-P-32 is a main sequence star, while HAT-P-67 has recently undergone reinflation and is either at the terminal stage of the main sequence or has evolved into a subgiant \citep{gully-santiago_large_2024}. Consequently, HAT-P-67~b currently receives approximately twice the incident flux compared to what it would have received at the zero age main sequence \citep{zhou_hat-p-67b_2017}. This enhanced irradiation may enable a stronger atmospheric escape from the planet’s dayside compared to HAT-P-32~b.

% what are other observations of pre-transit observations?
While a few other gas giants also show pre-transit absorption, none are as far-reaching as those of HAT-P-67~b and HAT-P-32~b. For example, observations and models of WASP-12~b suggest early-ingress absorption, indicative of metals overflowing the Roche lobe \citep{fossati_detailed_2010, llama_shocking_2011, haswell_near-ultraviolet_2012, jensen_hydrogen_2018}. However, \cite{nichols_hubble_2015} found no evidence of early ingress in the near-UV transit, contradicting earlier findings, although they did observe significant variability in the near-UV count rate before the optical transit. Surprisingly, helium was not detected in WASP-12~b, as reported by \cite{kreidberg_2018} and \cite{czesla_elusive_2024-1}. \cite{2018haex.bookE..97H} suggests that the outflow might form a torus around the star, making it difficult to detect excess absorption. Additionally, \cite{czesla_elusive_2024-1} observed variability in the stellar H-$\alpha$ and He-1083~nm lines across pre- to post-transit phases in the stellar rest frame, which may be the result of stellar activity. The structure of the outflow from WASP-12~b remains unclear. Its low surface gravity ($\log g = 3.0$) and extremely short orbital period of 1.1 days, suggest that the planet is a good candidate to form a cold planetary wind. However, interpreting the helium and hydrogen signatures is complicated by stellar activity.

%--------------------------------------------------------------------
\section{Conclusion} \label{sec:concl}
We have demonstrated through our models that the characteristics of a leading outflow stream as observed through \hei observations in HAT-P-32~b and HAT-P-67~b (see Fig.~\ref{fig:HPF_obs}) are best explained by a high planetary mass-loss rate on the order of $\sim 5 \times 10^{12}$ to $\sim 1 \times 10^{13}$~g~s$^{-1}$ ($\sim 10^{-10} M_{\rm J}$~yr$^{-1}$) and a relatively cool outflow temperature of approximately $\sim 4700$~K, which is just a few times the equilibrium temperature, primarily originating from the heated dayside of the planet. 

Our models suggest that to create a stream with the majority of absorption occurring pre-transit, a relatively cool outflow is essential (see Fig. \ref{fig:IsoAniso}). A dayside-dominated outflow is favored over an isotropic outflow, as it shows greater asymmetry in the light curve (see Fig. \ref{fig:EW}). In contrast, the hot wind scenarios fail to replicate the extensive light curve of excess \hei absorption across the orbital phase.

The \hei line of the cold wind scenario shows a large absorption depth and an irregular line profile. Our analysis reveals that the line is not broadened thermally, but instead kinematically. The irregular shape arises from distinct velocity distributions within the outflow (Fig. \ref{fig:spectra}). By measuring both the EW and velocity shift over the orbital phase, it is possible to constrain both the outflow and its shaping by the stellar wind (Figs.~\ref{fig:EW} and \ref{fig:stellar_wind+shift}). 

\section*{Data availability}
We have uploaded the updated radiative transfer code together with simulation material on Zenodo (\href{https://doi.org/10.5281/zenodo.13988501}{10.5281/zenodo.13988501}).

\begin{acknowledgements}
      The authors acknowledge the Texas Advanced Computing Center (TACC) at the University of Texas at Austin for providing computational resources that have contributed to the research results reported within this paper (\url{http://www.tacc.utexas.edu}). Furthermore, we thank SURFsara (\url{www.surfsara.nl}) for their support in using the Lisa Compute Cluster. We appreciate the helpful discussions with D. Linssen, K. Baka, K. Lange, and C. Dominic. A. Oklop\v{c}i\'{c} gratefully acknowledges support from the Dutch Research Council NWO Veni grant.
\end{acknowledgements}
%--------------------------------------------------------------------
\bibliography{main}{}
\bibliographystyle{aa}
%--------------------------------------------------------------------

\appendix
\section{Simulation movie}
\label{sec:appA}
\begin{figure}[h!]
    \centering
    \includegraphics[width=0.5\textwidth]{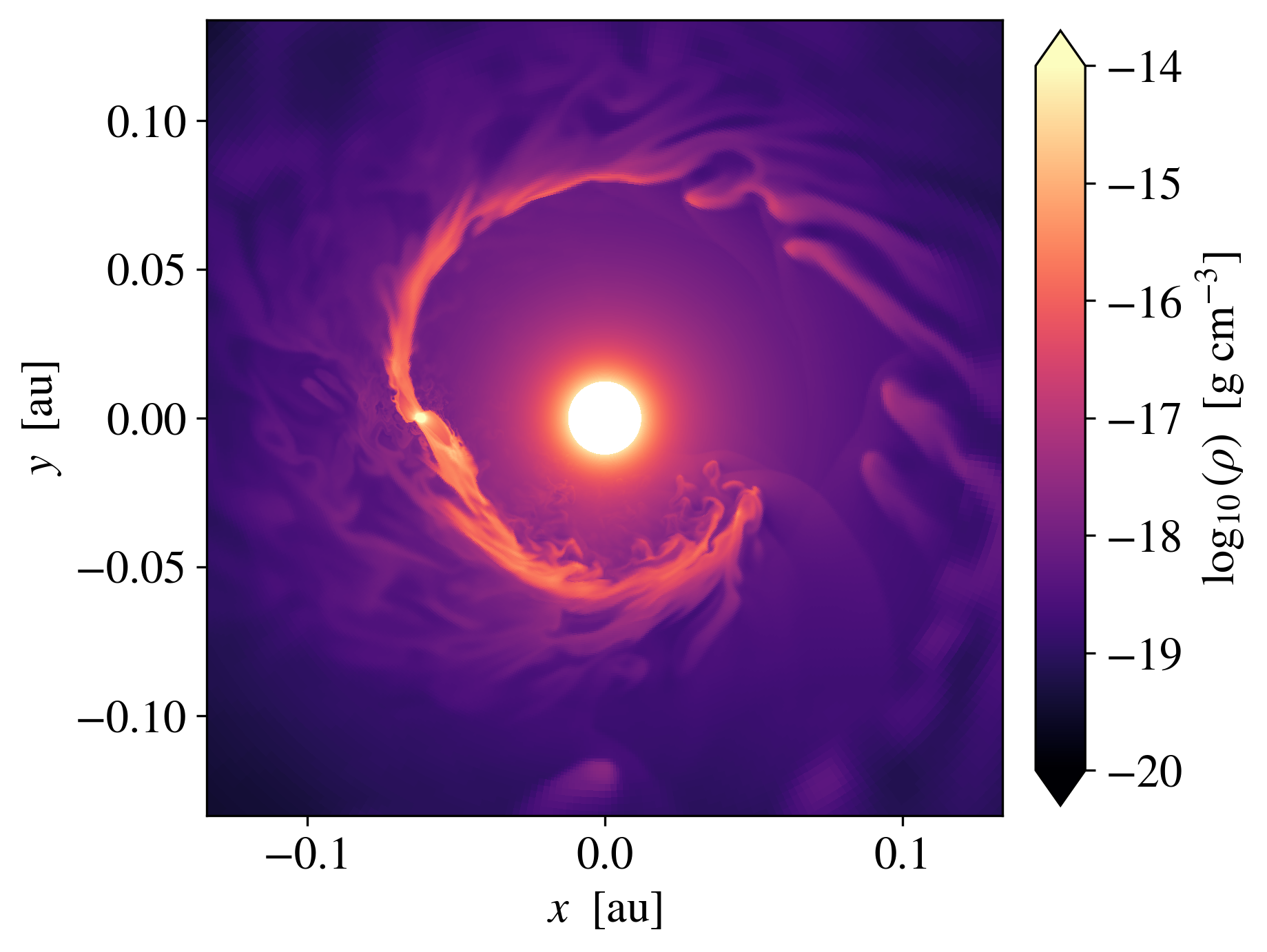}
    \caption{Gas density in the orbital midplane from the 3D simulation of fiducial model S2 over one orbit. An associated movie is available online. }
    \label{fig:movie}
\end{figure}

The online movie associated with Fig.~\ref{fig:movie} presents a high-resolution time sequence of the fiducial model S2 over the course of one orbit.

\section{Line formation in streams}
\label{sec:line_formation}

In this section we provide a more detailed analysis of the formation of the \hei line and the three peaks observed in the spectrum, as indicated in Fig.~\ref{fig:spectra}. We focus on model S2*, which has a higher density scaling compared to model S2, resulting in more clearly defined peaks. For clarity, we examined only the main (red) component of the helium triplet, which consists of two unresolved spectral lines, but effectively behaves like a singlet line.

\begin{figure}
    \centering
    \includegraphics[width=0.5\textwidth]{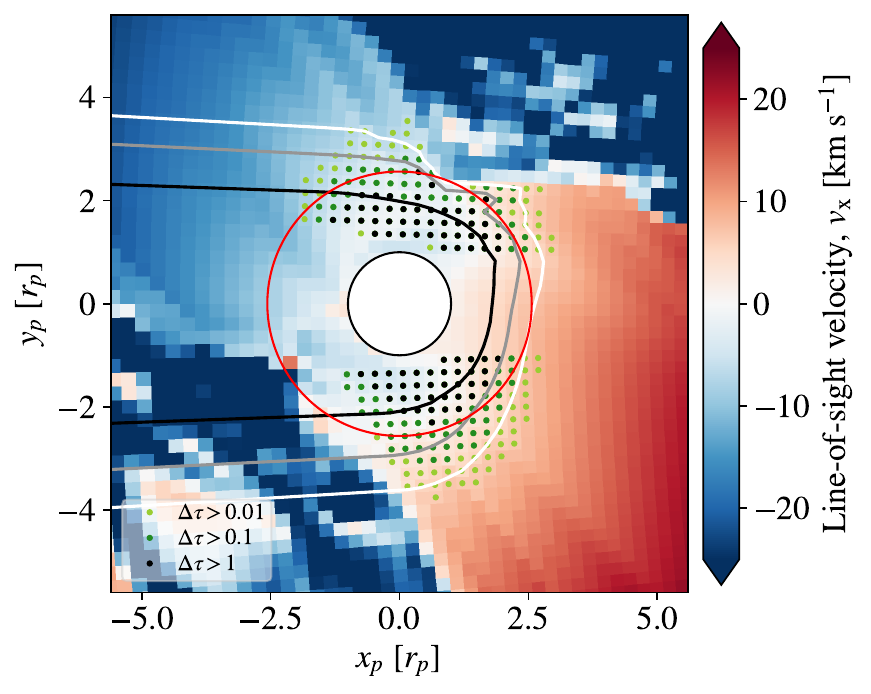}
    \caption{Line-of-sight velocity in the vicinity of the planet in the orbital midplane for simulation model S2*. The observer is positioned along the negative x-axis, with the velocity component in the x direction representing the observer's perspective during mid-transit. The black and red circles indicate the planetary radius and the Hill radius, respectively. This analysis focuses solely on the main component of the metastable helium line. Black, gray, and white contours mark the cumulative optical depth of $\tau = 10$, 1, and 0.1, respectively, along the radial direction toward the observer. The dots indicate high optical depth values ($\Delta \tau$) in the simulation grid. This figure highlights the primary region of \hei line formation. 
}
    \label{fig:v-map}
\end{figure}

Figure~\ref{fig:v-map} displays the line-of-sight velocity in the orbital midplane near the planet. The leading stream is directed away from the observer (red), while the trailing stream is moving toward the observer (blue). At the terminator regions, the gas velocity relative to the observer is approximately zero (white). High optical depth cells are highlighted to show the regions where most of the helium main component forms.

\begin{figure}
    \centering
    \includegraphics[width=0.45\textwidth]{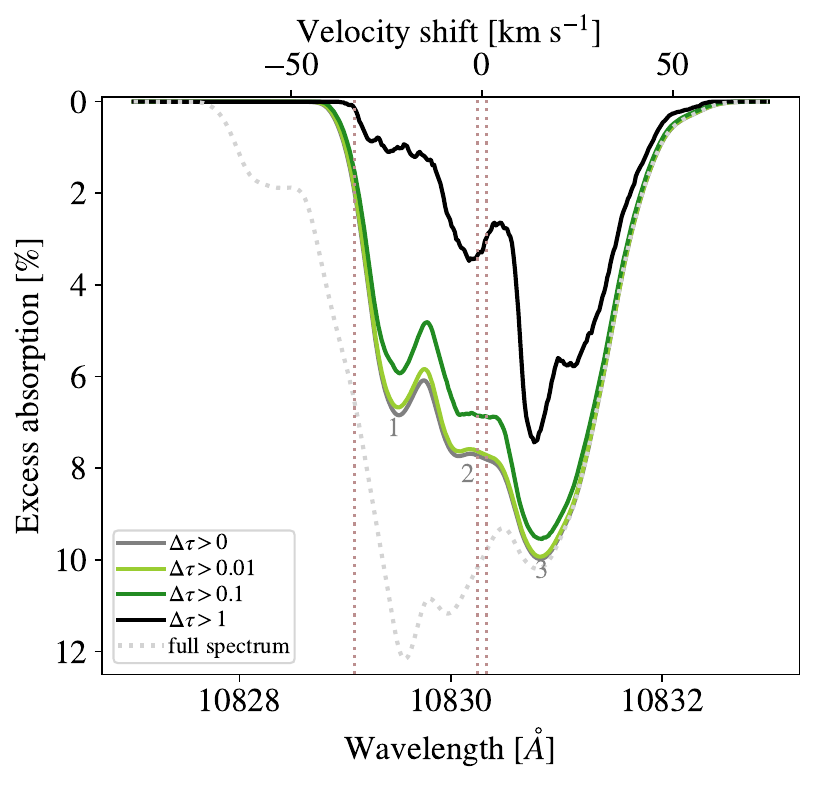}
    \caption{Red component of the metastable helium triplet of model S2*, calculated from cells that exceed a specific threshold in optical depth (see Fig.~\ref{fig:v-map}). The velocity axis is centered around the peak of the red component, which consists of two unresolved spectral lines, marked by the vertical dotted lines. The three peaks in the spectrum correspond to the trailing stream (1, blueshifted), the region near the planet (2, almost no shift), and the leading stream (3, redshifted). For completeness, we also include the full \hei spectrum, which incorporates the blue component of the triplet, shown with a dotted gray line.
}
    \label{fig:dtau_spectra}
\end{figure}

In Fig.~\ref{fig:dtau_spectra} the spectra were calculated with a threshold applied to the optical depth of the cells considered. The black spectrum includes only cells with optical depth greater than 1, while the green spectra gradually includes cells with progressively lower optical depth thresholds. The most significant change occurs when the threshold is lowered from $\Delta\tau > 1$ to $\Delta\tau > 0.1$, where the intermediate green cells indicated in Fig.~\ref{fig:v-map} contribute the most. The resulting spectrum reveals three distinct peaks: the first corresponds to the blue-shifted trailing stream, the second represents the terminator region of the planet with minimal velocity shift, and the third arises from the redshifted leading stream. Due to the day-night anisotropy, the leading stream has a higher density than the trailing stream, which is reflected in the depths of the peaks and the overall redshift of the spectrum. For completeness, we also include the full spectrum, which incorporates the blue component of the helium triplet. The three peaks remain visible in the full spectrum.

\end{document}